\documentclass[]{spie}  

\usepackage{amsmath,amsfonts,amssymb}
\usepackage{graphicx}
\usepackage[colorlinks=true, allcolors=blue]{hyperref}
\usepackage{siunitx}
\usepackage{textcomp, gensymb}
\usepackage{caption}
\usepackage{subcaption}

\DeclareSIUnit{\sr}{sr}

\title{Development of next-generation event-driven x-ray hybrid CMOS detectors}

\author[1]{Timothy R. Emeigh}
\author[1]{Abraham D. Falcone}
\author[1,2]{Joseph M. Colosimo}
\author[1]{Kadri M. Nizam}
\author[1]{Lukas R. Stone}
\author[1]{Md. Arman Hossen}
\author[1]{Laurel ONeill}
\author[1]{Ian Ashcroft}
\author[1]{Kyle Bonene}
\author[1]{Brynn Bortree}
\author[1]{Zachary E. Catlin}
\author[1]{Sierra Deppe}
\author[1]{Killian M. Gremling}
\author[1]{Gavin Houlihan}
\author[1]{Katie McWhirter}
\author[3]{David M. Palmer}
\author[1]{Abigail A. Raytsis}

\affil[1]{The Pennsylvania State University, Astronomy and Astrophysics, 525 Davey Lab, University Park, USA}
\affil[2]{now at NASA Goddard Space Flight Center, Greenbelt, MD, USA}
\affil[3]{Los Alamos National Laboratory/NMC, Los Alamos, NM, USA}

\authorinfo{Further author information:\\T.R.E.: E-mail: tre8@psu.edu}

\begin{document} 
\maketitle

\begin{abstract}
The Penn State University High Energy Astrophysics Detector and Instrumentation Lab—in collaboration with Teledyne Imaging Sensors—has developed a next-generation event-driven X-ray hybrid CMOS detector. This detector is an HCD with a 1,024$\times$1,024 array of 21-$\mu$m pitch pixels, featuring a comparator in each pixel to enable event-driven readout. While the full frame operation of this detector can reach readout speeds of up to 150 Hz, this event-driven operation allows readout of only those pixels that surpass a user-set charge threshold, enabling effective rates of up to 10 kHz. Here we report on the past detector development efforts at Penn State and describe these new detectors and expected performance improvements.
\end{abstract}



\section{Introduction}
\label{sect:introduction}
\begin{figure}[ht]
\centering
\begin{tabular}{c}
\centering
\includegraphics[width = 8 cm]{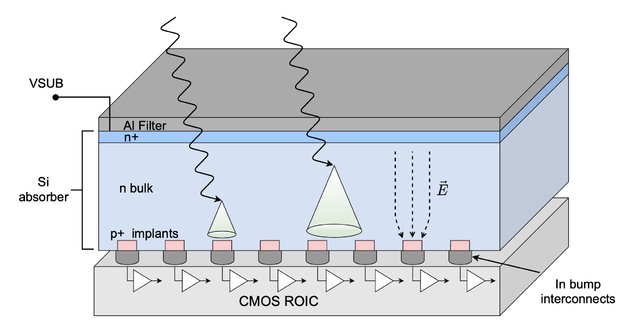}
\end{tabular}
\caption 
{ \label{fig:xhcd}
Render of an HCD comprising two layers: a ROIC wafer indium-bump-bonded to a high-resistivity silicon absorbing layer. This design enables separate optimization of each layer.} 
\end{figure}

X-ray hybrid CMOS detectors (HCDs) are active pixel sensors comprising two separate silicon substrates hybridized together through indium bump bonds. This affords both substrates individual characteristics best suited to their task. The bottom substrate is the readout integrated circuit (ROIC) optimized for efficient, fast, and low-noise readout. The top substrate is a high-resistivity silicon absorbing layer optimized for full depletion to depths of 100 $\mu$m, allowing improved quantum efficiency across the soft X-ray band (up to 20 keV).\cite{Bongiorno2015, Colosimo2022, Arman2026} Fig.~\ref{fig:xhcd} shows a rendering of a generalized X-ray HCD.

X-ray HCDs offer several advantages over the charge-coupled devices (CCDs) typically in use by current X-ray observatories. Due to the CMOS active pixel sensors (APS) design, the ROIC is inherently low power at much faster readout rates than CCDs.\cite{Kozlowski2002, Falcone2021} Additionally, the APS and back-illuminated design provides increased radiation and impact resiliency, isolating any damage that does occur to singular pixels and reducing the likelihood of such damage in the first place by placing the more resilient absorber bulk ahead of the sensitive readout electronics.\cite{Bray2020} Each of these factors remain critical for large-effective-area, high-throughput future \textit{Lynx}-like observatories, with fast frame rates avoiding pile-up and enabling deeper time-domain studies, and lower power draw and resiliency allowing the longevity needed for these missions.\cite{Falcone2019} While these missions do also demand low read noise, and HCDs typically face worse performance in that regard than the best science-grade CCDs, this is an area of active development.

Penn State University (PSU) has developed a next-generation X-ray HCD in collaboration with Teledyne Imaging Sensors (TIS). This device features a 1,024$\times$1,204 array of pixels at a 21-$\mu$m pitch, with global-shutter full frame readout reaching speeds of 150 Hz. Each pixel has a high-gain capacitive transimpedance amplifier, in-pixel correlated double sampling, and comparator enabling event-driven readout with effective readout rates of up to 10 kHz.

\section{Preceding Detectors}
\label{sect:oldets}
As completed devices have not yet been received by Penn State for in-lab testing and initial characterization results, we present here the results from preceding detector development efforts at Penn State, with Sect.~\ref{sect:nextgen} delivering expectations from these upcoming devices as a culmination of this past work.

As HCDs are active-pixel sensors, with each pixel having its own circuitry and thus performance, read noise values are quoted here as the mode of the distribution of per-pixel RMS read noises. Energy resolutions are quote as the FWHM of the best fit of the single-pixel event spectra for each device. Unless otherwise specified, characterization of each device was carried out under vacuum at a temperature of 150 K. Characteristic Mn K$\alpha$ X-rays were provided by a radioactive $^{55}$Fe source. All other energies were provided by an e- impact source or fluorescence of a target by incident radiation from a $^{210}$Po source.

\subsection{HxRG}
The HxRG family of HCDs are designs based on the TIS HAWAII hybrid Si CMOS detectors, with the "x" in the name standing in for an integer multiple of 1,024 as the pixel count along one side of the square ROIC array.\cite{Loose2003} The HxRGs require the Teledyne SIDECAR$^{\text{TM}}$ application-specific integrated circuit (ASIC) to provide clock and bias signals, charge amplification, and analog-to-digital conversion and readout. While PSU has previously investigated the performance of this ASIC at room temperature\cite{Griffith2012, Prieskorn2013}, it is designed as a cryogenic device and was used as such for all results presented here. This decision was made to suppress temperature-induced fluctuations in the electronics and produce the lowest read noise and energy resolution possible.

\begin{figure}[ht]
\centering
\begin{tabular}{c}
\centering
\includegraphics[width = 10 cm]{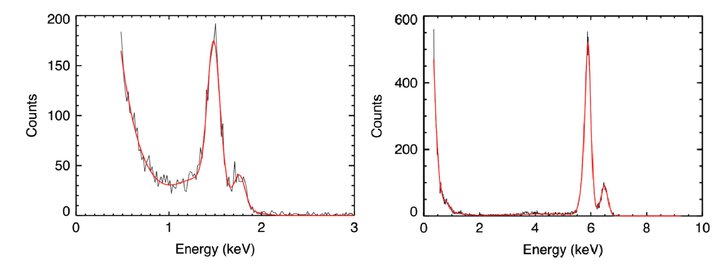}
\end{tabular}
\caption 
{ \label{fig:h1rgspect}
150 K H1RG single-pixel event spectra for Al K$\alpha$ and Mn K$\alpha$ respectively.}
\end{figure}

The first X-ray HCD was the H1RG, developed in 2006. It has a 1,024$\times$1,024 pixel array in both ROIC and absorber, with an 18-$\mu$m pitch. It also demonstrated the successful deposition of an Al optical blocking filter (OBF) directly onto the silicon absorbing layer.\cite{Falcone2007} The best measured read noise for the H1RG was 7.05 e- and the best energy resolution was 4.8\% (280 eV) at the Mn K$\alpha$ 5.9 keV line and 17.7\% (265 eV) at the Al K$\alpha$ 1.5 keV line.\cite{Prieskorn2013} The spectra used for these energy resolution measurements are shown in Fig.~\ref{fig:h1rgspect}.

Even though the RMS read noise is relatively low, these energy resolution values are well above expected theoretical values for these detectors. This is due to the presence of inter-pixel capacitance (IPC) in the H1RG.\cite{Finger2006} Due to the small pixel size and source follower amplifiers used in these devices, when charge is collected on the input node of a pixel, it can become capacitively coupled with neighboring pixels and artificially spread. This effect was found to a degree of 4.0–5.5\% charge sharing vertically and 8.7–9.7\% horizontally in the H1RG.\cite{Prieskorn2013}

The H2RG was fabricated with the elimination of this IPC being a key driving factor. The H2RG uses a ROIC array of 2,048$\times$2,048, 18-$\mu$m pitch pixels. The absorber array, however, is 1,024$\times$10,24 at a 36-$\mu$m pitch. This gives four ROIC pixels for each absorber pixel, with each of the latter only being bump bonded to one of those four, providing a buffer of inactive pixels between each active pixel. The increased physical separation of active pixels was indeed shown to reduce IPC, with measurements only finding $\sim$1.8\% charge sharing.\cite{Griffith2012}

\begin{figure}[ht]
\centering
\begin{tabular}{c}
\centering
\includegraphics[width = 10 cm]{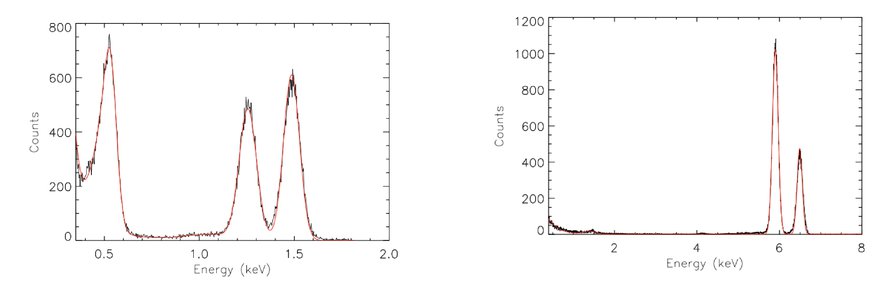}
\end{tabular}
\caption 
{ \label{fig:h2rgspect}
150 K H2RG single-pixel event spectra for O K$\alpha$ plus Al K$\alpha$ and Mn K$\alpha$ respectively.}
\end{figure}

Performance was also found to improve. The best measured read noise was slightly lower than in the H1RG at 6.8 e-. The energy resolution faced a much more dramatic improvement, reaching 2.7\% (160 eV) at Mn K$\alpha$ and 6.8\% (100 eV) at Al K$\alpha$. This improvement also allowed the measurement of the 0.53 keV O K$\alpha$ line for the first time in an HCD as it became well-separated from the noise peak, found to be 17.7\% (93.8 eV).\cite{Hull2017} The spectra used for measuring the energy resolution of the H2RG can be seen in Fig.~\ref{fig:h2rgspect}.

\subsection{Speedster-EXD}
The next detector produced in the collaboration between PSU and TIS is the Speedster-EXD detector, with both a 128$\times$128 pixel prototype and a larger 550$\times$550 pixel version. Both versions have a 40 $\mu$m pixel pitch. The Speedster-EXD introduced several changes to the readout circuitry: the source follower amplifier was replaced with a capacitive transimpedence amplifier (CTIA) to eliminate IPC\cite{Griffith2016}, in-pixel correlated double sampling was introduced to remove reset baseline variations, and a comparator was included to enable event driven readout (discussed further in Sect.~\ref{subsect:eventdriven}). Additionally, the Speedster-EXD is a digital-in, digital-out device. Digitization is performed with on-chip column-parallel single-slope ADCs, eliminating the need for the SIDECAR$^{\text{TM}}$ ASIC.

\begin{figure}[ht]
\centering
\begin{tabular}{c}
\centering
\includegraphics[width = 10 cm]{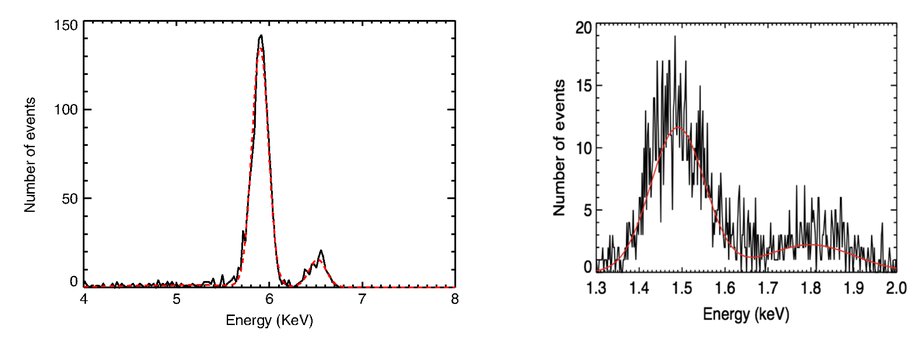}
\end{tabular}
\caption 
{ \label{fig:speedspect}
150 K Speedster-EXD single-pixel event spectra for Mn K$\alpha$ and Al K$\alpha$ respectively.}
\end{figure}

The best measured read noise achieved by a Speedster-EXD detector is 11.2e-. The best measured energy resolution is 3.5\% (210 eV) at Mn K$\alpha$ and 10\% (150 eV) at Al K$\alpha$.\cite{Griffith2016} Plots of the energy resolution can be seen in Fig.~\ref{fig:speedspect}. This further speaks to the elimination of IPC; despite a worse read noise than in the H1RG, the recovered energy resolution is still better.

\subsection{Event Driven Readout}
\label{subsect:eventdriven}
\begin{figure}[ht]
\centering
\begin{tabular}{c}
\centering
\includegraphics[width = 8.5 cm]{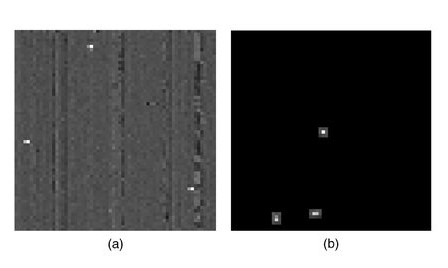}
\end{tabular}
\caption 
{ \label{fig:eventdriven}
Full frame and 3$\times$3 event driven readout of a Speedster-EXD detector, respectively.}
\end{figure}

While operating in full frame mode, reading out every pixel every frame, the Speedster-EXD is capable of 152 Hz readout rate. However, each pixel contains a comparator which, when provided a globally-set reference voltage, only triggers readout if that pixel's readout signal exceeds the reference. Readout can either be only that pixel (1$\times$1 mode) or that pixel and the eight adjacent pixels (3$\times$3 readout). With the column-parallel ADCs and limited digital data rate serving as bottlenecks to operation speeds, this increases the effective readout rate of the devices, reducing the amount of data being sent through those bottlenecks by only considering potential events, not dark pixels.\cite{Griffith2016} A comparison between full frame readout and 3$\times$3 event driven readout can be seen in Fig.~\ref{fig:eventdriven}

While operating in event driven readout, effective readout rates of up to 10 kHz can be reached. However, there are two detrimental impacts on the performance of the detectors: when the comparators are powered there is a mild induced kickback on the rest of the readout circuitry and the removal of dark pixels prevents measurement of frame-to-frame bias variations. These additional sources of noise worsen the energy resolution, with the best measured energy resolution in 3$\times$3 event driven mode for a Speedster-EXD is 4.2\% (250 eV).\cite{Griffith2016}

\subsection{BlackCAT CubeSat}
Despite the lessened performance compared to earlier devices, the 550$\times$550 Speedster-EXD (Speedster-EXD550) was selected for the Black Hole Coded Aperture Telescope (BlackCAT) CubeSat mission and deemed still capable of meeting the mission's primary and secondary science goals. The BlackCAT mission is a wide-FOV (0.85 sr partially coded) X-ray transient detector and X-ray sky monitor,\cite{Falcone2024} launched in January 2026. The focal plane array consists of our four best Speedster-EXD550 detectors, making these devices the first X-ray HCDs flown on an orbital mission and increasing the Speedster-EXD550 TRL to 9.

In preparation for this mission, which is passively cooled to 233 K (-40$\degree$C) with a radiator panel\cite{Wages2025}, calibration was needed at these higher temperatures. Additionally, considerations to settings were made to optimize for a balance between read noise and energy resolution performance and overall power draw. With these considerations made, the focal plane array has a measured read noise of approximately 20e- (RMS) and an energy resolution of 4.2\% (250 eV) in full frame operation at BlackCAT temperatures.\cite{Emeigh2025}.

Additionally, while characterizing these detectors for BlackCAT, the presence of random telegraph noise (RTN) was identified on the order of 10\% of pixels, where the baseline voltage levels can randomly shift between at least two default states. In detectors with CDS like the Speedster-EXDs, RTN presents itself as a triple- or multi-peaked distribution of dark pixel values, depending on if and between which states the voltage shifted between the two samples of CDS. For shifts between voltage states that are at a small enough amplitude, this increases the relevant pixel's noise. For larger shifts, this can significantly reduce the recovered charge of an event or cause spurious triggers on false events without a true X-ray deposited. Events from pixels with high frequency or amplitude RTN will be removed from the flight readout, increasing bad pixel fraction.\cite{Colosimo2024}

\begin{figure}[ht]
\centering
\begin{tabular}{c}
\centering
\includegraphics[width = 11.5 cm]{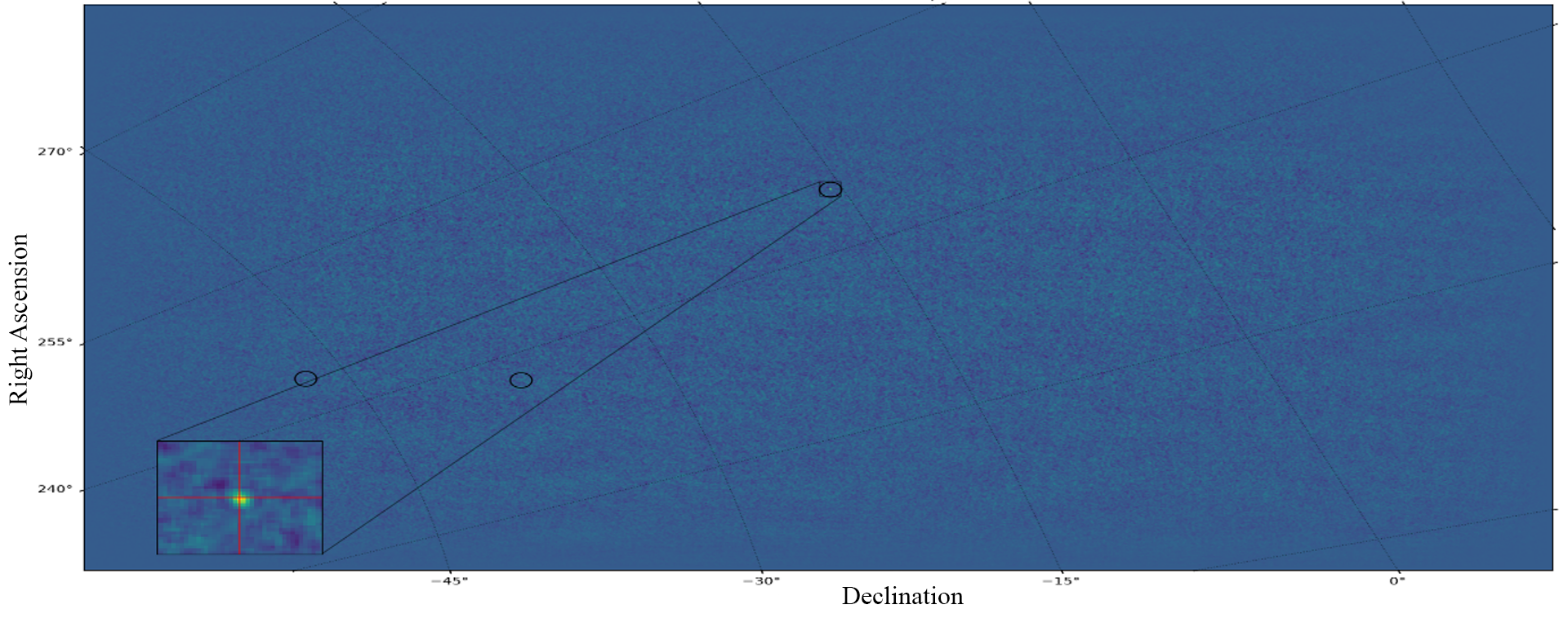}
\end{tabular}
\caption 
{ \label{fig:scox1}
Reconstructed image from events detected by Speedster-EXD550 detectors onboard BlackCAT. The inset shows the localized peak associated with Scorpius X-1.}
\end{figure}

Despite these shortcomings, the Speedster-EXD550s on BlackCAT have proven operational and capable in-flight. During the payload commissioning phase, astronomical X-rays were detected by the devices as a first light. Shortly afterward, Scorpius X-1 moved into our accessible field of view, providing a bright source to test image reconstruction and source localization capabilities in-flight. Fig.~\ref{fig:scox1} shows such a reconstructed image, with an inset zooming in on the location of Sco X-1. Additionally, there are two peaks at lower significance, which are circled. These represent other potential detections in this early-commissioning-phase observation.

\subsection{Small-Pixel Detectors}
\begin{figure}[ht]
\centering
\begin{tabular}{c}
\centering
\includegraphics[width = 10 cm]{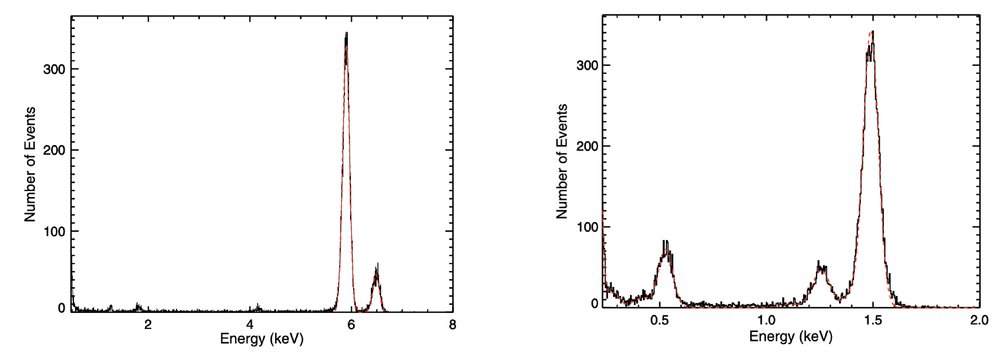}
\end{tabular}
\caption 
{ \label{fig:smallspect} Single-pixel event spectra for Mn K$\alpha$ on the left and O K$\alpha$, Mg K$\alpha$, and Al K$\alpha$ on the right.
}
\end{figure}

The next detectors developed by PSU and TIS are the Small-Pixel HCDs, with 12.5 $\mu$m pixel pitch in a prototype 128$\times$128 array and final 1,024$\times$1,024 array. These devices use the same approach of a CTIA and in-pixel CDS as the Speedster-EXD HCDs, but with the specifics of pixel design modified, a different provider for silicon wafers, and no comparator and subsequently no event driven readout. These changes have culminated in a design that provides the best measured read noise and energy resolutions from an HCD to date. The best Small-Pixel HCD read noise was measured at 5.4 e-. The best measured energy resolutions were 2.7\% (160 eV) at Mn K$\alpha$, 6.1\% (91 eV) at Al K$\alpha$, 7.9\% (99 eV) at the 1.25 keV Mg K$\alpha$ line, and 15\% (78 eV) at O K$\alpha$.\cite{Hull2019} These results can be seen in Fig.~\ref{fig:smallspect}.

\section{Next Generation X-Ray HCDs}
\label{sect:nextgen}
\begin{figure}[ht]
\centering
\begin{tabular}{c}
\centering
\includegraphics[width = 10 cm]{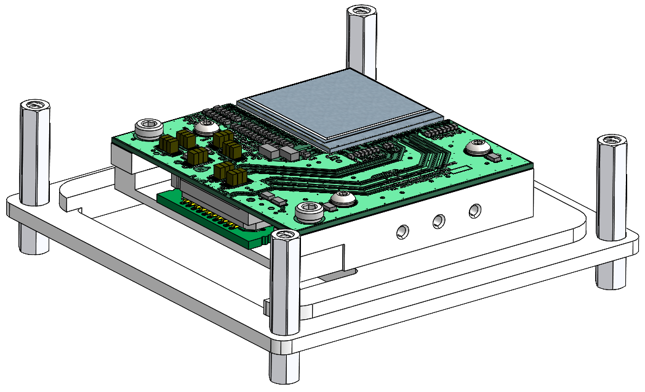}
\end{tabular}
\caption 
{ \label{fig:nextgen} CAD model of the next-generation X-ray HCD.
}
\end{figure}

PSU and TIS have developed a next-generation X-ray HCD and are in the progress of producing the initial engineering units of this device. This device features an array of 1,024$\times$1,024 pixels at 21-$\mu$m pitch, with global-shutter full frame readout reaching speeds of 150 Hz. Each pixel is based on a simplified version of the Small-Pixel HCD design with the addition of an in-pixel comparator to enable event-driven readout, enabling effective readout speeds of up to 10 kHz. Silicon wafers for production were also sourced from the same manufacturer as in the Small-Pixel HCD. We thus expect a device with the read noise, energy resolution, and lower RTN (compared to the Speedster-EXD) of the Small-Pixel HCD while having the event-driven capabilities of the Speedster-EXD. A CAD model of the next-generation X-ray HCD can be seen in Fig.~\ref{fig:nextgen}.

Beyond just selecting the best of both worlds of the two previous HCDs developed, these next generation detectors offer several fixes and improvements to bugs and flaws seen in these older devices, as well as several new features.

In the Speedster-EXD HCDs, functionality was included to deselect pixels from event driven readout to avoid excess triggers from faulty comparators, hot circuitry, or readout noise. This functionality only worked in $\sim$50\% of columns, however.\cite{Colosimo2024}. Additionally, setting the ROI to exclude a row was supposed to prevent any readout of that row, regardless of if events were present or not. However, due to a bug, if an event fell in an ROI row and tripped a compartor, that row would be read out as blank instead of not read out at all. Both of these flaws would contribute to reduced effective readout rates in event driven mode, and both have been addressed in the next-generation X-ray HCD design.

In the Small-Pixel HCDs, two different charge-injection schemes were tested, designated A0 and A1. Both of these charge-injection schemes were found to be too strong, pinching off the CTIA and leading to low-energy gain nonlinearity in the CTIA. While the A0 charge-injection can be turned off, this leads to a loss of effective full well instead, modestly impacting X-ray bandpass.\cite{Stone2026} To address this, the charge-injection in the next-generation X-ray HCD has been designed with several, user-selectable states, allowing it to be optimized to balance between linearity and X-ray bandpass.

Two additional changes were made to the device which are original to these next-generation detectors. The first of these is an increase to the maximum voltage acceptable before the on-chip decoupling capacitors trip. In the Speedster-EXD HCDs this was 25 V, while in the Small-Pixel HCDs this was 100 V; it has been increased to 200 V in these new devices. Given the same substrate thickness of 100 $\mu$m as these past two devices, the larger pixels than on the Small-Pixel, and the significantly higher substrate voltage, we expect reduced charge spreading and a larger fraction of single pixel events, reducing the impact of event grading on overall energy resolution when considering all good photon events.

Finally, the engineering model next-generation X-ray HCDs are being produced with two 64-row strips of increased gain. The first of these strips increases the pixel gain by $\sim$33\%, while the second increases it by $\sim$90\%. We expect this change to add minimal additional electronics noise while simultaneously lessening the conversion of what noise there is to its electron equivalent. This will ultimately lead to a lowered readout noise in these detectors even compared to the Small-Pixel HCDs they are based on, with a target readout noise of <4 e- in these strips. While this does come at the cost of pixel full well, future production models of these detectors will be able to use any of the three circuit designs for the entire array based on use case and the results of calibration on the initial engineering run.

\section{Conclusion}
\label{sect:conclusion}
We have presented the history of X-ray HCD development at PSU in collaboration with TIS, culminating in a presentation of the next-generation event-driven X-ray HCD and its expected performance. While these devices have not yet been delivered to PSU for confirmation, we anticipate similar performance to the Small-Pixel HCDs at $\sim$5.5 e- read noise, with the possibility of pushing lower to $\leq$4 e- in specialized high-gain strips of pixels. Additionally, bugs found in the previous Speedster-EXD and Small-Pixel circuit design have been addressed and the maximum substrate voltage has been improved to 200 V in these new detectors, further driving improved performance in secondary measures beyond read noise and energy resolution, such as charge spreading, bad pixel masking, and event-driven readout rates.

\acknowledgments
We would like to acknowledge our collaborators at Teledyne Imaging Sensors for their contributions to this project. This material is based upon work supported by the National Aeronautics and Space Administration under
Agreement No. 80NSSC24K0327 and 80NSSC21k1125 issued through the Office of Science. In accordance with Federal law, NMC is prohibited from discriminating on the basis of race, color, national origin, sex, age, or disability.


\bibliography{report}   
\bibliographystyle{spiebib} 

\end{document}